\documentclass[conference]{IEEEtran}
\usepackage{cite}
\usepackage{soul} 
\usepackage{xcolor}
\usepackage{upgreek}

\usepackage[cmex10]{amsmath}
\usepackage{centernot}
\usepackage{array}
\usepackage{url}

\usepackage{graphicx}
\usepackage{amsmath} 
\usepackage{cite}

\newenvironment{example}{\par\begin{quotation}\small\noindent{\bf Example:\ }}{\end{quotation}\par}

	\newcommand{\twodef}[4]{\left\{\begin{array}{ll} 
		\displaystyle {#1} & {#2} \\
		\displaystyle {#3} & {#4}
		\end{array}\right.}

\newcommand{\blankout}[1]{}

\newcommand{\be}{\begin{equation}}
\newcommand{\ee}{\end{equation}}
\newcommand{\bee}{\begin{IEEEeqnarray}{c}}
\newcommand{\eee}{\end{IEEEeqnarray}}
\newcommand{\bs}{\begin{IEEEeqnarray*}{c}}
\newcommand{\es}{\end{IEEEeqnarray*}}
\newcommand{\bd}{\begin{IEEEeqnarray*}{c}}
\newcommand{\ed}{\end{IEEEeqnarray*}}
\newcommand{\bea}{\begin{IEEEeqnarray}{rCl}}
\newcommand{\eea}{\end{IEEEeqnarray}}
\newcommand{\bas}{\begin{IEEEeqnarray*}{rCl}}
\newcommand{\eas}{\end{IEEEeqnarray*}}
\newcommand{\equat}[1]{equation (\ref{eq:#1})}
\newcommand{\Equat}[1]{Equation (\ref{eq:#1})}

\newcommand{\Ns}{{x_S}}
\newcommand{\dotNs}{{\dot{x}_S}}
\newcommand{\Nsorder}{{x_S^{\order(t)}}}

\newcommand{\Nsstar}{{x_S^*}}
\newcommand{\Ni}{{x_I}}
\newcommand{\dotNi}{{\dot{x}_I}}
\newcommand{\Nr}{{x_R}}

\newcommand{\NIt}{{{\cal X}_I}}

\newcommand{\fe}[1]{{f_{{\cal E}{(#1)}}}}
\newcommand{\fee}{f_{{\cal E}}}
\newcommand{\Fe}[1]{{F_{{\cal E}{(#1)}}}}
\newcommand{\Epbar}{{\bar{\cal E}}}

\newcommand{\Eppbar}{{\overline{{\cal E}^2}}}
\newcommand{\Epppbar}{{\overline{{\cal E}^3}}}
\newcommand{\dEpbar}{{{\dot{\Epbar}}}}

\newcommand{\Rep}{{\cal R}_{\epsilon}}
\newcommand{\epmax}{{\epsilon_{\mbox{\tiny  max}}}}
\newcommand{\epz}{{\epsilon_0}}
\newcommand{\dt}{{\mbox{\small $\Delta$} t\,}}
\newcommand{\sgamma}{{\theta}}
\newcommand{\Ep}{{\cal E}}
\newcommand{\Epprime}{{\Ep^\dagger}}
\newcommand{\Epbarprime}{{{\Epbar}^\dagger}}
\newcommand{\Epsig}{{{\sigma_{\Ep}^2}}}
\newcommand{\order}{{\cal K}}
\newcommand{\dorder}{{\dot{\order}}}
\newcommand{\Kw}{{\cal W}}

\newcommand{\dEpsig}{{\dot{\Epsig}}}
\newcommand{\dNIt}{{\dot{\NIt}}}
\newcommand{\dNit}{{\dot{\Ni}}}
\newcommand{\dNst}{{\dot{\Ns}}}
\newcommand{\Epskew}{{{\cal S}_{\Ep}^3}}
\newcommand{\boe}{{herd immunity threshold}}
\newcommand{\Boe}{{Herd Immunity Threshold}}

\author{Christopher Rose,$^{1*}$
Andrew J. Medford,$^2$
C. Franklin Goldsmith,$^1$ 
Tejs Vegge,$^3$\\
Joshua S. Weitz,$^{4*}$
Andrew A. Peterson$^{(1,3)*}$\\
\normalsize{$^1$School of Engineering, Brown University, Providence, Rhode Island, 02912, USA}\\
\normalsize{$^2$School of Chemical \& Biomolecular Engineering, Georgia Institute of Technology, Atlanta, Georgia, 30332, USA}\\
\normalsize{$^3$Department of Energy Conversion and Storage, Technical University of Denmark, 2800 Kgs.~Lyngby, Denmark}\\
\normalsize{$^4$School of Biological Sciences and School of Physics, Georgia Institute of Technology, Atlanta, Georgia, 30332, USA}\\
\normalsize{$^*$Corresponding authors: jsweitz@gatech.edu, andrew\_peterson@brown.edu, christopher\_rose@brown.edu}
}

\title{Population Susceptibility Variation and Its Effect on  Contagion Dynamics}

\begin{document}

\maketitle

\begin{abstract}
  Susceptibility governs the dynamics of contagion. The classical SIR model is one of the simplest
  compartmental models of contagion spread, assuming a single shared susceptibility level.  However,
  variation in susceptibility over a population can fundamentally alter the dynamics of contagion
  and thus the ultimate outcome of a pandemic.  We develop mathematical machinery which explicitly
  considers susceptibility variation, illuminates how the susceptibility distribution is sculpted by
  contagion, and thence how such variation affects the SIR differential questions that govern
  contagion. Our methods allow us to derive closed form expressions for herd immunity thresholds as
  a function of initial susceptibility distributions and suggests an intuitively satisfying approach
  to inoculation when only a fraction of the population is accessible to such intervention.  Of
  particular interest, if we assume static susceptibility of individuals in the susceptible pool,
  ignoring susceptibility diversity {\em always} results in overestimation of the herd immunity
  threshold and that difference can be dramatic. Therefore, we should develop robust measures of
  susceptibility variation as part of public health strategies for handling pandemics.
\end{abstract}


\section{Introduction}

The differential equations typically used to describe contagion \cite{Brauer2012,Corlan2015} place the population into three different tranches:
\begin{itemize}
    \item $\Ns$: susceptible fraction/number 
    \item $\Ni$: infected fraction/number
    \item $\Nr$: recovered fraction/number
\end{itemize}
Assuming the fractional form, we have
$$
\Ns + \Ni + \Nr = 1
$$
There are also two key parameters governing contagion dynamics:
\begin{itemize}
    \item $\beta$: the rate  ($(\mbox{individual-time})^{-1}$) of transmission
    \item $\gamma$: the rate ($\mbox{time}^{-1}$) of recovery 
\end{itemize}
which lead to the fundamental coupled differential equations of contagion
$$
{\dotNs}
=
- \beta \Ns \Ni
$$
and
$$
{\dotNi}
=
\left (\beta \Ns - \gamma \right ) \Ni
$$

However, all members of a population are not necessarily as susceptible to contagion as others \cite{Dwyer1997,Kwiatkowski2000, Lloyd-Smith2005}.  So, let $\epsilon \ge 0$ be the susceptibility of an individual to a given disease.  Small values of $\epsilon$ imply greater resistance, while large values imply greater susceptibility.  We can then define the random variable ${\cal E}(t)$ as the susceptibility of an individual chosen randomly from the susceptible population at time $t$.  Its probability density function is
$\fe{t}(\epsilon)$ and
$$
\Fe{t}(\epsilon) = \int_0^{\epsilon} \fe{t}(x) d x
$$
is its cumulative distribution function -- the probability that an individual randomly selected from the population at time $t$ will have a susceptibility less than or equal to $\epsilon$.

Now consider a Gedankenexperiment where individuals are selected randomly from the population and exposed
to contagion. Our key assumption is that
\begin{quote}
{\bf \boldmath Individuals with susceptibility $\epsilon$ will be removed from the susceptible pool at a rate $\beta {\Ni} \epsilon$.}
\end{quote}

Over time, such removals will alter the population susceptibility landscape $\fe{t}(\epsilon)$.
That is, individuals with higher susceptibility are preferentially removed early, and this process, repeated
many times, will increase the relative proportion of less susceptible individuals.  We seek to understand
in general how $\fe{t}(\epsilon)$ evolves in time.  So we amend the equations of contagion as
\be
\label{eq:dNsdt}
{\dNst}
=
- \beta \Epbar(t) \Ns \Ni
\ee
and
\be
\label{eq:dNidt}
{\dNit}
=
\left ( \beta \Epbar(t) \Ns - \gamma \right )
\Ni
\ee
where $\Epbar(t)$ is the mean susceptibility of the population -- which we take to be initially $\Epbar(0) = 1$.

We will find that if the initial susceptibility distribution, $\fe{0}(\epsilon)$ is Gamma-distributed, then $\fe{t}(\epsilon)$ stays Gamma-distributed.  
However, we also find that the contagion
process, if left to run long enough, tends to sculpt $\fe{0}(\epsilon)$ into an approximation of
a Gamma distribution.  The exceptions include initial mixed (singular $+$ continuous) distributions as well as those with non-compact support.  However, if the initial distribution can be expressed over its domain as a power series (including series representations with non-integer powers), then $\fe{t}(\epsilon)$ approaches a Gamma distribution of some order.  But most importantly, given the general assumptions of the SIR model and assuming static individual susceptibility we will find that 
\begin{itemize}
\item
{\bf  Ignoring susceptibility diversity {\em always} results in overestimation of the herd immunity threshold.}
\item
  {\bf  The population susceptibility distribution {\em shape} affects}
  \begin{itemize}
\item
  {\bf the ultimate severity of contagion.}
\item
  {\bf the effectiveness of mitigation techniques}    
\end{itemize}
\end{itemize}

\section{Evolution of $\fe{t}(\epsilon)$}
Taking a differential approach consider that for a small time-step $\dt$,
the probability density $\fe{t+\dt}(\epsilon)$ must be
\bas
\fe{t+\dt}(\epsilon) &  = &
\frac{\displaystyle \fe{t}(\epsilon) (1 - \beta {\Ni}(t + \dt)  \dt  \epsilon )}{\displaystyle \int \fe{t}(\epsilon) \left (1- \beta {\Ni}(t+\dt) \dt \epsilon \right ) d\epsilon} \\
 & = & \frac{ ( 1 - \beta {\Ni}(t+\dt) \dt \epsilon) }{1 - \beta {\Ni}(t+\dt) \dt  \Epbar(t)}\fe{t}(\epsilon)
\eas
We then have $\frac{\fe{t + \dt}(\epsilon) - \fe{t}(\epsilon)}{\dt}$ as
$$
\frac{\beta {\Ni}(t+\dt) \dt(\Epbar(t) - \epsilon )}{ \dt \left (1 - \beta {\Ni}(t+\dt) \dt  \Epbar(t) \right )}
\fe{t}(\epsilon) 
$$
which after $\dt$ disappears from numerator and denominator leaves
$$
\frac{\beta {\Ni}(t+\dt) 
(\Epbar(t) - \epsilon)}{1 - \beta {\Ni}(t+\dt) \dt \Epbar(t)} \fe{t}(\epsilon)
$$
which as $\dt \rightarrow 0$ reduces to
\be
\label{eq:evolution}
\boxed{
\frac{d}{dt} \fe{t}(\epsilon)
=
\beta {\Ni} ( \Epbar(t) - \epsilon) \fe{t}(\epsilon)
}
\ee
\Equat{evolution} is the differential equation governing the evolution of $\fe{t}(\epsilon)$ in time under the action of contagion.  We immediately see that  susceptibility above average will be muted while susceptibility below average will be amplified. As \equat{evolution} evolves, we will expect $\Epbar(t)$ to decrease and the probability mass of $\Ep(t)$ to become more and more concentrated around smaller values of susceptibility.

\subsection{A General Solution}
We assume one individual's susceptibility does not affect another's.  So, we can imagine a given susceptibility tranche as being exponentially diminished according to its susceptibility value $\epsilon$. If $\Ns (\epsilon,0)$ is the size of that tranche at time zero, then we may expect
\be
\label{eq:tranchedecay}
\Ns(\epsilon,t) \propto  e^{-\beta {\NIt} \epsilon} \Ns(\epsilon,0)
\ee
where we define
\be
\label{eq:infectionpressure}
{\NIt} = \int_0^t {\Ni} dt
\ee
as the cumulative "infection pressure."
We note that so long as ${\Ni}$ does not contain singularities, ${\NIt}(0) = 0$.  We also note that ${\NIt}(t)$ is non-negative and non-decreasing -- and not necessarily bounded if individuals neither recover nor die.

Thus, if $g_0(\epsilon)$ is the initial distribution of susceptibility at time zero, we posit
\be
\label{eq:general}
\boxed{
\fe{t}(\epsilon)
=
\frac{\displaystyle g_0(\epsilon) e^{-\beta {\NIt} \epsilon}}{\displaystyle \int g_0(\epsilon) e^{-\beta {\NIt} \epsilon} d \epsilon }
}
\ee
Checking for satisfaction of \equat{evolution}, we have $\frac{d}{dt} \fe{t}(\epsilon)$ as
$$
\left (
\!- \beta {\Ni} \epsilon  +\!\! \frac{\displaystyle \int \!\!\beta {\Ni} \epsilon g_0(\epsilon) e^{-\beta {\NIt} \epsilon} d \epsilon}{\displaystyle \int \!\!g_0(\epsilon) e^{-\beta {\NIt} \epsilon} d \epsilon}
\right )
\!\!\frac{\displaystyle g_0(\epsilon) e^{-\beta {\NIt} \epsilon}}{\displaystyle \int\!\! g_0(\epsilon) e^{-\beta {\NIt} \epsilon} d \epsilon }
$$
which we rewrite as
$$
\left (
\!- \beta {\Ni} \epsilon  + \!\! \int\!\! \beta {\Ni} \epsilon \frac{g_0(\epsilon) e^{-\beta {\NIt} \epsilon}}{\displaystyle \int \!\!g_0(\epsilon) e^{-\beta {\NIt} \epsilon} d \epsilon} d \epsilon \!\!
\right )
\!\!\frac{\displaystyle g_0(\epsilon) e^{-\beta {\NIt} \epsilon}}{\displaystyle \int\!\! g_0(\epsilon) e^{-\beta {\NIt} \epsilon} d \epsilon }
$$
which reduces to
$$
\frac{d}{dt}
\fe{t}(\epsilon)
 =
\beta {\Ni} \left ( \Epbar(t) - \epsilon \right ) \fe{t}(\epsilon)
$$
as required by \equat{evolution}.  So, \equat{general} is the general solution \cite{simmons} to the first order homogeneous linear differential \equat{evolution}.

\subsection{Susceptibility Distribution Evolution Examples}
\noindent{\bf \boldmath $2$-Point ${\fee}(\epsilon)$:} 
Suppose
\be
\label{eq:2point}
{\fee}(\epsilon)
=
p\delta(\epsilon) +  (1-p) \delta\left (\epsilon - \frac{\Epbar}{1-p}  \right )
\ee
with mean $\Epbar$ and variance $\frac{p}{1-p} \Epbar^2$.  Application of \equat{general} yields
\be
\label{eq:f2point}
\fe{t}(\epsilon)
=
\frac{p\delta(\epsilon) +  (1-p)  (1-p) e^{-\frac{\beta\NIt \Epbar}{1-p}}
\delta\left (\epsilon - \frac{\Epbar}{1-p}  \right )}
{p\delta(\epsilon) +  (1-p) e^{-\frac{\beta\NIt \Epbar}{1-p}}}
\ee

\vspace{0.125in}
\noindent{{\boldmath \bf Uniform $g_0(\epsilon)$:}}
Suppose
\be
\label{eq:gunif}
g_0(\epsilon)
=
\frac{1}{\epmax}
(u(\epsilon) - u(\epsilon - \epmax))
\ee
Using \equat{general} we obtain
\be
\label{eq:funiform}
\boxed{
\fe{t}(\epsilon)
=
\frac{\beta {\NIt}}{1 - e^{-\beta {\NIt} \epmax}} e^{-\beta {\NIt} \epsilon }
}
\ee
for $\epsilon \in [0, \epmax]$.
So, as the cumulative number of infections grows, $\fe{t}(\epsilon)$ becomes exponential on the
interval $[0,\epsilon]$.  As ${\NIt}$ grows, the mean susceptibility time course for this distribution approaches
$$
\Epbar(t) \stackrel{\mbox{$\NIt$ large}}{\longrightarrow} \frac{1}{\beta {\NIt}}
$$
The exact time course of $\Epbar(t)$ is given by
\be
\label{eq:epbaruniform}
\boxed{
\Epbar(t)
=
\frac{1}{\beta \NIt} \cdot \frac{1 - \left( 1 + \beta \NIt \epmax\right) \cdot e^{-\beta \NIt \epmax}}{1 - e^{-\beta \NIt \epmax}}
}
\ee

\vspace{0.125in}
\noindent{{\boldmath \bf Gamma-distributed $g_0(\epsilon)$}:}
Suppose
\be
\label{eq:ggamma}
g_0 (\epsilon) =
\frac{\epsilon^{k-1} e^{-\epsilon k / \Epbar_0}}{\left( \frac{\Epbar_0}{k} \right)^k \Gamma(k)}
\ee
where $k$ is the shape parameter of the distribution, $\Epbar_0$ is the initial mean susceptibility
and $\Gamma(k)$ is the gamma function.  We see that as $k \rightarrow \infty$, the distribution
becomes an impulse at the mean.

Using \equat{general} we obtain
\be
\label{eq:fgamma}
\boxed{
\fe{t}(\epsilon)
=
\frac{\left(\frac{k}{\Epbar_0} + \beta {\NIt}\right)^k \epsilon^{k-1} e^{-\left(\frac{k}{\bar{\epsilon}_0} + \beta {\NIt} \right) \epsilon} }
{\Gamma(k)}}
\ee
which is itself a gamma distribution of order $k$ with mean susceptibility time course
\be
\label{eq:epbargamma}
\boxed{
\Epbar(t)=
\frac{1}{\frac{\beta {\NIt}}{k} + \frac{1}{\Epbar_0} }
}
\ee
We note that when $k \to \infty$, the $\Epbar(t)$ does not change with time -- a Gamma distribution
approaches an impulse at the mean for large $k$.  We also note that Gamma distributions appear to be
a sort of "eigenfunction" of the transformation on $g_0(\epsilon)$ applied by \equat{general}.
Specifically, if $g_0(\epsilon)$ is a Gamma function of order $k$, then $\fe{t}(\epsilon)$ is also a
Gamma function of order $k$ as seen in \equat{fgamma} with mean given by \equat{epbargamma}.

\vspace{0.125in}
\noindent {{\boldmath \bf Pareto-distributed $g_0(\epsilon)$}:}
Suppose
\be
\label{eq:pareto}
g_0(\epsilon)
=
\alpha \epz^\alpha  \epsilon^{-(1+\alpha)}
\ee
where $\alpha >2$ and  $\epsilon \ge \epz > 0$. Using \equat{general} we obtain
$$
\int_{\epz}^{\infty}
g_0(\epsilon) e^{-\beta {\NIt} \epsilon} d \epsilon
=
\alpha E_{(1+\alpha)}(\beta {\NIt} \epz)
$$
where
$$
E_n(z)
=
\int_1^\infty \frac{e^{-zx}}{x^n} dx
$$
so that
\be
\label{eq:fpareto}
\boxed{
\fe{t}(\epsilon)
=
\frac{\alpha \epz^\alpha  \epsilon^{-(1+\alpha)}}
{\alpha E_{(1+\alpha)}(\beta {\NIt} \epz)} e^{- \beta {\NIt} \epsilon}}
\ee
The mean susceptibility time course is given by
\be
\label{eq:epbarpareto}
\boxed{
\Epbar(t)
=
\epz \frac{E_{(\alpha)}(  \NIt \beta \epz)}{E_{(1 + \alpha)}( 
 \NIt \beta \epz)}
}
\ee
which in the limit of large $\NIt \beta$ approaches $\epz$ -- as is expected since the action of contagion (\equat{evolution}) drives $\Epbar(t)$ toward its absolute minimum, which in the case of a Pareto distribution, is $\epz$.

\subsection{Rate of Mean Susceptibility Change}
The change in the average susceptibility as a function of time for any given distribution $\fe{t}(\epsilon)$ is:
\bas
\frac{d}{dt} \Epbar(t) & = & \frac{d}{dt} \int \epsilon \fe{t}(\epsilon)  d \epsilon \\
 & = & \int \beta \dNIt ( \epsilon \Epbar(t) - \epsilon^2) \fe{t}(\epsilon) d \epsilon \\
\eas
which reduces to
\be
\label{eq:dEdt}
\boxed{
\frac{d}{dt} \Epbar(t) 
=
- \beta \dNIt \sigma_{{\cal E}(t)}^2
= - \beta \Ni \sigma_{{\cal E}(t)}^2
}
\ee
where $\sigma_{{\cal E}(t)}^2 $ is the variance of $\Epbar(t)$. Since the infection pressure $\NIt$
is non-decreasing, $\frac{d}{dt} \Epbar(t)\le 0$ -- as expected since contagion preferentially
removes the more susceptible.

\subsection{Contagion Sculpts the Susceptibility Density}
\label{sect:sculpt}
\begin{figure*}[ht]
\begin{center}
\begin{tabular}{cc}
\includegraphics[width=3.25in]{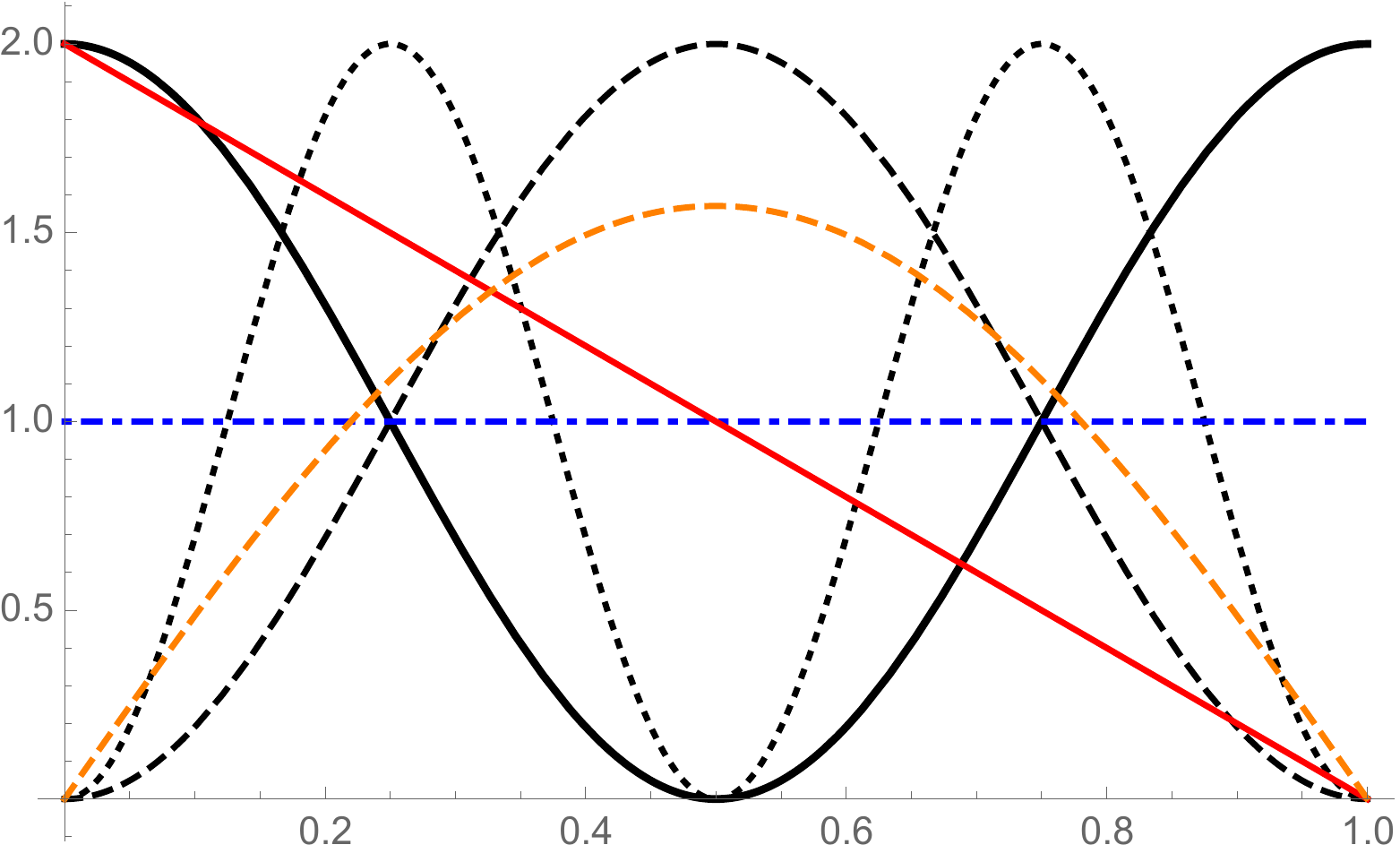} & \includegraphics[width=3.25in]{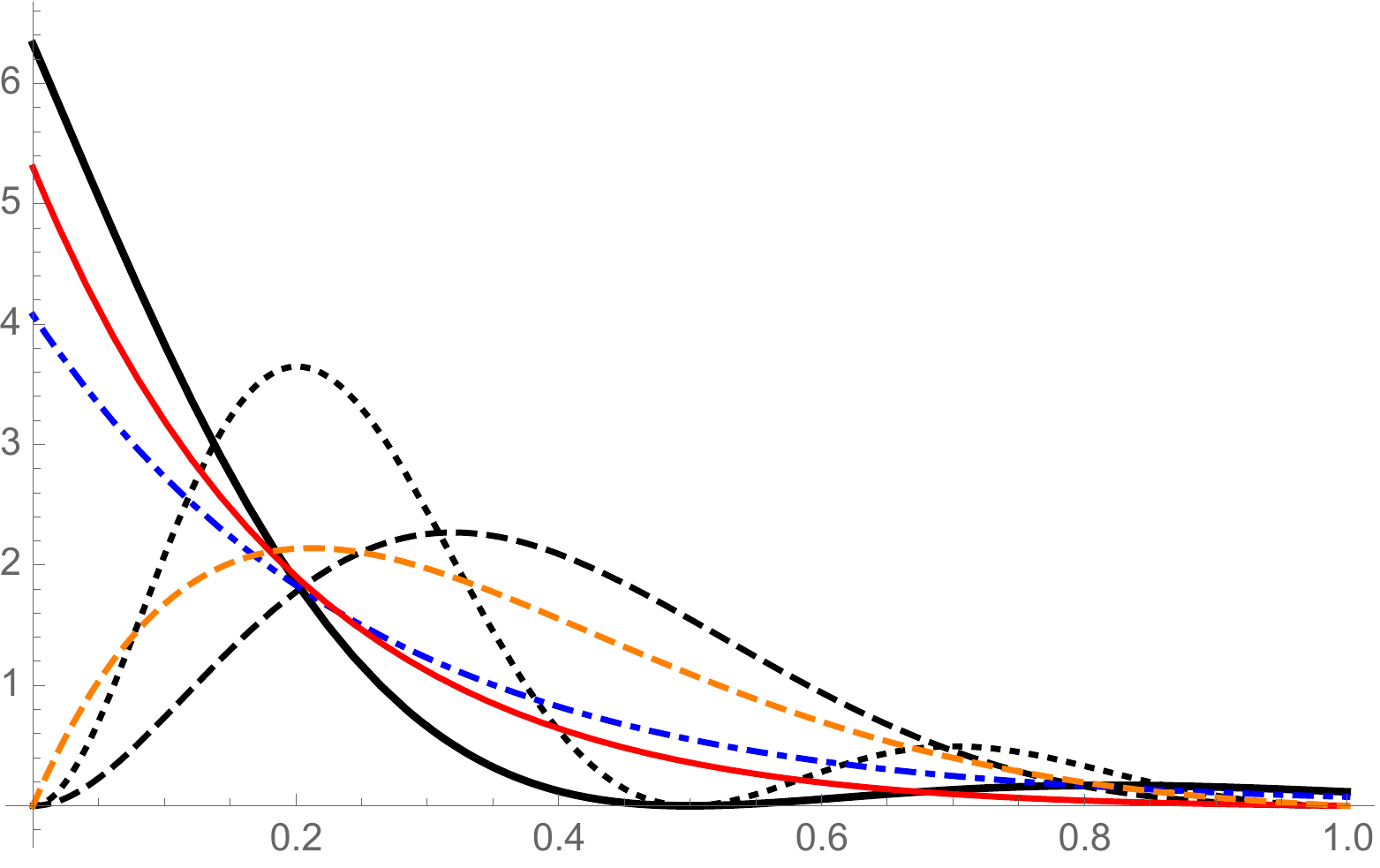} \\
{\Large \bf (a)} & {\Large \bf (b)}\\
$\mbox{}$ & $\mbox{}$\\
\includegraphics[width=3.25in]{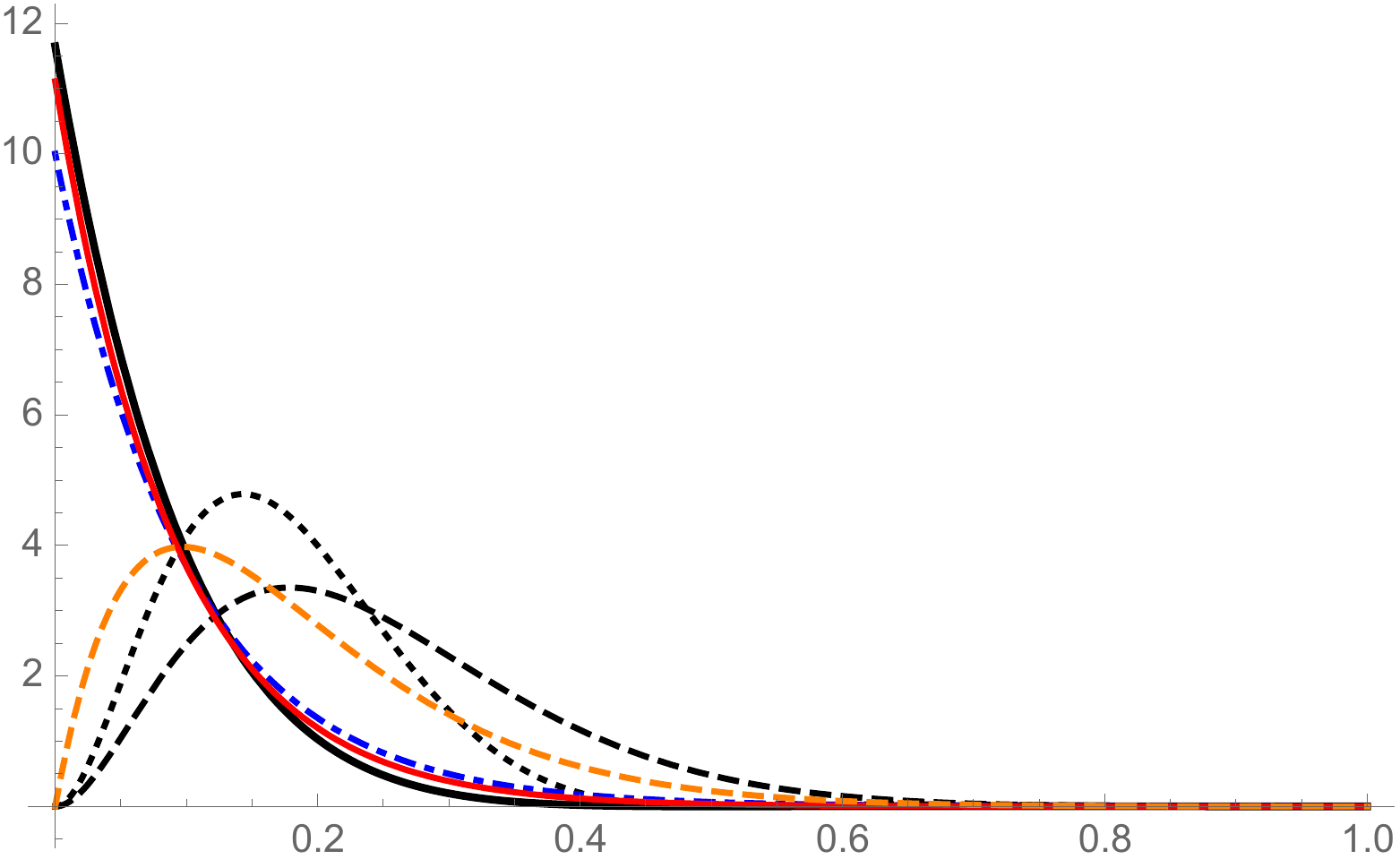} & \includegraphics[width=3.25in]{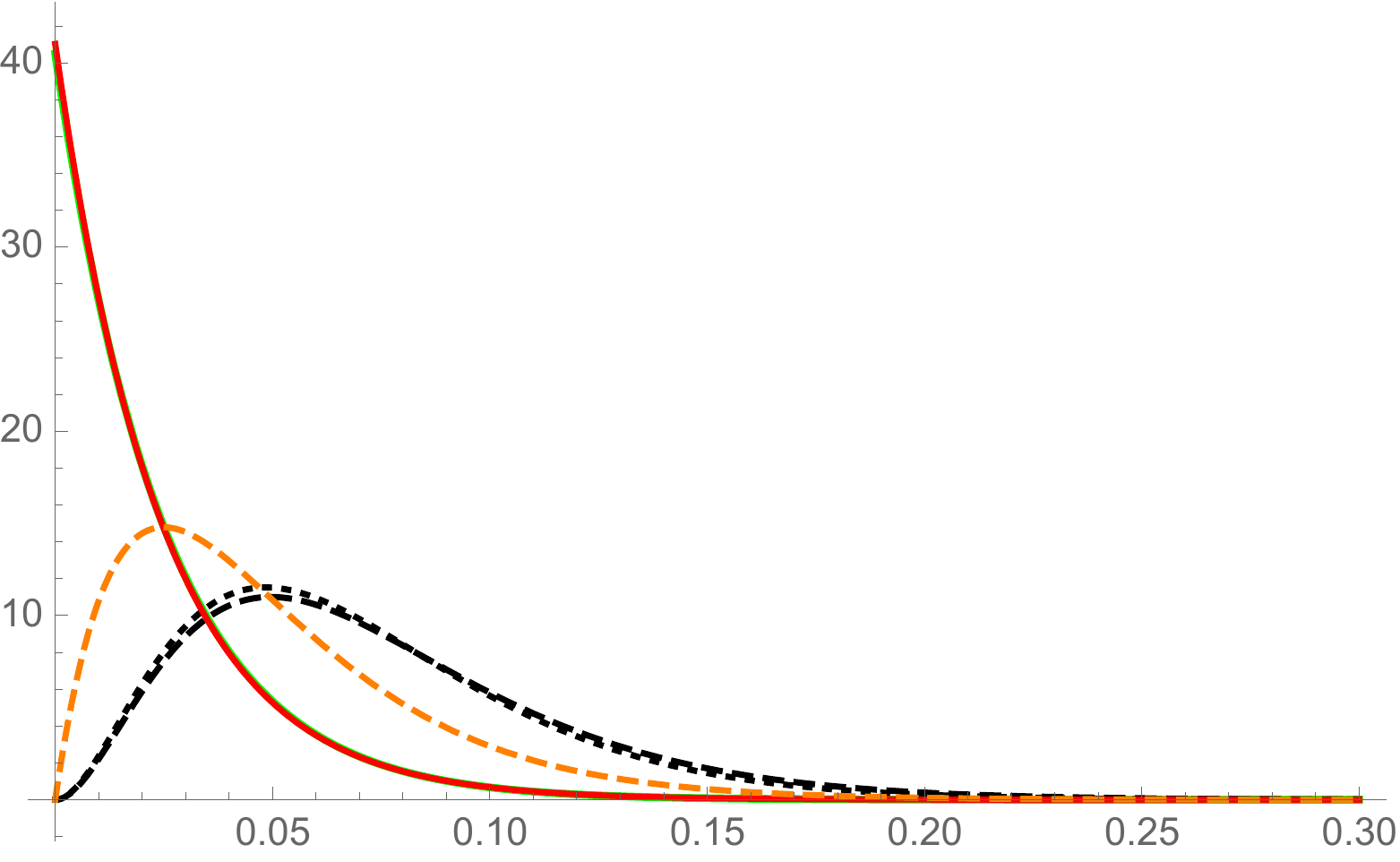}\\
{\Large \bf (c)} & {\Large \bf (d)}\\
\end{tabular}
\end{center}
\caption{ {\bf \boldmath Sculpting of $\fe{\beta \NIt}(\epsilon)$ with $\beta {\NIt}$:} {
    Solid -- $g_0(\epsilon) = 1 + \cos 2 \pi \epsilon$; Dashed --
    $g_0(\epsilon) = 1 - \cos 2 \pi \epsilon$; Dotted -- $g_0(\epsilon) = 1 - \cos 4 \pi \epsilon$;
    Dot-Dashed -- $g_0(\epsilon)$ uniform; Red -- $g_0(\epsilon)$ downsloped linear; Orange --
    $g_0(\epsilon) = \frac{\pi}{2} \sin \pi \epsilon$.  Note convergence to exponential (Gamma
    distribution with $k=1$) for those initial distributions that are approximately constant for
    small $\epsilon$, Gamma with $k=2$ for the lone sinusoidal distribution that is proportional to
    $\epsilon$ for small $\epsilon$, and Gamma with $k=3$ for the two cosinusoidal distributions with $g_0(0) = 0$ which are quadratic for small $\epsilon$.
    {\bf (a)} $\beta {\NIt} = 0$; {\bf (b)} $\beta {\NIt} =4$;
    {\bf (c)} $\beta {\NIt} = 10$; {\bf (d)} $\beta {\NIt} = 40$.} }
\label{fig:FGang}
\end{figure*}
It is easy to see that Gamma distributions are a sort of ``eigenfunction'' for susceptibility
distribution evolution -- an initially Gamma $g_0(\epsilon)$ guarantees that $\fe{t}(\epsilon)$ will
remain Gamma of the same order as $g_0(\epsilon)$ $\forall t$.  This raises the possibility that the
Gamma distribution is an {\em attractor} of \equat{evolution}.  This is not so.  However, if
$g_0(\epsilon)$ is continuous and can be expressed as a power series (Taylor/MacLauren or
fractional), then we can show that $\fe{t}(\epsilon)$ will indeed approach a Gamma distribution.

First, define the compact region $\Rep$ with boundaries $\epsilon^- < \epsilon^+$,
$$
\Rep
=
\{ \epsilon | \epsilon^- \le \epsilon  \le \epsilon^+ \}
$$
such that
$$
\int_{\epsilon^-}^{\epsilon^+}  \fe{t}(\epsilon) d \epsilon \approx 1
$$
If $\exists \Rep$ such that $g_0(\epsilon)$ is well-approximated for $\epsilon \in \Rep$ by some $\theta \epsilon^\ell$ with $\ell > -1$ and $\theta > 0$, then 
\be
\label{eq:sculpt}
\fe{t}(\epsilon)
\approx
C(\theta,\beta\NIt) \theta \epsilon^\ell e^{-\beta \NIt \epsilon}
\ee
where $C(\sgamma,\beta\NIt)$ is an appropriate normalization constant.
\Equat{sculpt} is a Gamma distribution of order $k = \ell+1$, mean $\frac{k}{\beta \NIt}$ and variance $\frac{k}{(\beta\NIt)^2}$.  If $k$ is an integer the distribution is also Erlang (as well as Gamma), and if $k=1$ the distribution is exponential.

Then, note that \equat{evolution} dictates the probability mass of $\fe{t}(\epsilon)$ will be forced
to the left with increasing time (indexed by $\NIt$) because $e^{-\beta \NIt \epsilon}$ necessarily
concentrates the probability mass (and thereby the region $\Rep$) closer to the origin.  Thus, even
if $g_0(\epsilon)$ is arbitrary in form away from the origin, so long as $\Rep$ eventually covers a
region where $g_0(\epsilon)$ approximately $\theta \epsilon^\ell$ for some $\ell>-1$, the density
$\fe{t}(\epsilon)$ will eventually be approximately a Gamma distribution with parameter
$k = \ell+1$.  The evolution of several different initial distributions is shown in
FIGURE~\ref{fig:FGang}.  The convergence to Gamma distributions of orders $k=1,2,3$ is as expected
from small-$\epsilon$ order of the initial distributions -- first order in the three cases where
$g_0(0) \ne 0$, second order for the lone sinusoidal distribution which is linear for small
$\epsilon$, and third order for the two cosinusoidal distributions with $g_0(0) = 0$ which are
quadratic for small $\epsilon$.

This argument can also be extended to cases were $\ell \le -1$ so long as for $\epsilon^- > 0$ we
have $g_0(\epsilon) = 0$ $\forall \epsilon < \epsilon^-$.  To our knowledge the resultant
distribution, \equat{sculpt} with $\ell \le -1$, has no formal name.  However, we note that Pareto
$g_0(\epsilon)$ will produce $\fe{t}(\epsilon)$ in this $\ell \le 1$ class.
The situation is of course more complicated if $g_0(\epsilon)$ cannot eventually be
well-approximated by $\theta \epsilon^\ell$ on some limiting $\Rep$, is without compact support or
contains singularities.

\section{$\Epbar(t)$ and the Number of Susceptibles, $\Ns(t)$}
Typically, $\Epbar$ is considered independent of population variables when evaluating the
differential equations of contagion.  However, when the distribution on susceptibility in a
population is {\em not} singular, we will show -- generalizing the development in
\cite{Peterson2020} -- that $\Epbar$ can depend strongly on the number of susceptible individuals, $\Ns(t)$,
according to the susceptibility distribution, $\fe{t}(\epsilon)$.

For notational clarity we will drop the time variable $t$, recognizing that all quantities are
functions of time under the action of contagion, including the distribution on susceptibility. Thus, 
if $\Ns$ is the total number of susceptible individuals in a population at time $t$, we assume the number, $n_S(\epsilon)$, of individuals with susceptibility $\epsilon$ is
$$
n_S(\epsilon) = \Ns {\fee}(\epsilon)
$$
We can then define $E$ as the average susceptible population (as opposed to the average susceptibility of individuals, $\Epbar$) as
\be
\label{eq:E}
E = \int_{0}^{\infty}\epsilon n_S(\epsilon) d \epsilon 
=
\Ns  \Epbar
\ee

Now, define the random variable $\Epprime$ as the susceptibility of those who have just fallen ill at time $t$.  The distribution of $\Epprime$ is
\be
\label{eq:Epprime}
f_{\Epprime}(\epsilon)
=
\frac{\epsilon}{\Epbar} {\fee}(\epsilon)
\ee
and it has mean
\be
\label{eq:Epbarprime}
\Epbarprime
=
\int \frac{\epsilon^2}{\Epbar} {\fee}(\epsilon) d \epsilon
=
\frac{\Epsig}{\Epbar} + \Epbar
\ee
where $\Epsig$ is the variance of the susceptibility at time $t$.  But we can also interpret $\Epbarprime$ as the rate of change of the total susceptibility $E$ with respect to $\Ns$.  That is, let $\Delta\!\Ns$ be the differential number of individuals {\em removed} from the susceptible pool during a time instant $\dt$.
Since the newly infected's susceptibilities follow the distribution of \equat{Epprime}, the {\em decline} $\Delta\!E$ in $E$ is
$$
\Delta\! E
=
\Epbarprime(t) \Delta\! \Ns
$$
and the ratio of $\Delta\!E$ to $\Delta\! \Ns$ as $\dt \rightarrow 0$ is
\be
\label{eq:dEdN}
\frac{dE}{d \Ns} 
=
\Epbarprime
\ee

Now, differentiating \equat{E} with respect to $\Ns$ yields
$$
\frac{d E}{d \Ns}
=
\Epbar + \Ns \frac{d \Epbar}{d\Ns}
$$
which through application of \equat{dEdN} becomes
$$
\Epbarprime
=
\Epbar + \Ns \frac{d \Epbar}{d\Ns}
$$
which via \equat{Epbarprime} simplifies to
\be
\label{eq:NSandEpbarPRE}
\frac{\Epsig}{\Epbar} 
=
\Ns \frac{d \Epbar}{d\Ns}
\ee
which we rearrange as
$$
\frac{d \Ns}{\Ns}
=
d \Epbar \left ( {\frac{\Epbar}{\Epsig}} \right )
$$
so that assuming $\Epbar(0) =1$ we have
\be
\label{eq:NSandEpbar}
\boxed{
\log \left ( \frac{\Ns}{\Ns(0)} \right )
=
\int_1^{\Epbar}  \left (\frac{\Epbar}{\Epsig} \right )  d \Epbar
}
\ee

\Equat{NSandEpbar} tells us that $\Epbar$ is explicitly a function of the contagion state variable
$\Ns$, a dependence which {\em fundamentally subverts the assumption of average susceptibility
  as an independent parameter} in the contagion dynamical equations.  Rather $\Epbar(t)$ is a
contagion {\em state variable}.  Put another way, the dependence of $\Epbar$ on $\Ns$ changes the
order of the contagion differential equations, and this order may in fact be a function of time. 

In the next section we explore \equat{NSandEpbar} for several different susceptibility distribution
types to motivate more formally defining an instantaneous order, $\order$, of $\Epbar$ with respect
to $\Ns$.

\subsection{$\Ns$ vs. $\Epbar$ Examples}
The key element of \equat{NSandEpbar} is the expression $\frac{\Epbar}{\Epsig}$ and its dependence on $\Epbar$. For any given distribution, $\Epsig$ and $\Epbar$ may be independent or dependent.  For instance, the mean and variance of a Gaussian distribution are independent -- one can be changed without affecting the other. In contrast, the variance of an exponential distribution is the square of the mean.  For many distributions, however, the mean and variance are neither as separable nor as crisply dependent, so to evaluate the integral of \equat{NSandEpbar} we must carefully find $\frac{\Epbar}{\Epsig}$ as a function of $\Epbar$ (and
other quantities independent of $\Epbar$).

\vspace{0.125in}
\noindent{\bf \boldmath $2$-Point ${\fee}(\epsilon)$:} 
Suppose
$$
{\fee}(\epsilon)
=
p\delta(\epsilon) +  (1-p) \delta\left (\epsilon - \frac{\Epbar}{1-p}  \right )
$$
with mean $\Epbar$ and variance $\frac{p}{1-p} \Epbar^2$. Thus
$$
\frac{\Epbar}{\Epsig}
=
\frac{1-p}{p \Epbar}
$$
Notice that selection of $p$ does not affect the mean, $\Epbar$, so we can safely apply
\equat{NSandEpbar} to obtain
\be
\boxed{
\label{eq:S2point}
\Epbar =  \left (\frac{\Ns}{\Ns(0)} \right )^{\frac{p}{1-p}}
}_{\,\,\, \mbox{$2$-Point}}
\ee

\vspace{0.125in}
\noindent{\bf \boldmath Uniform ${\fee}(\epsilon)$:} 
Suppose ${\fee}(\epsilon)$ is uniform on $[\epsilon^-, \epsilon^+]$.  We then have
$$
\Epbar = \frac{\epsilon^+ - \epsilon^-}{2}
$$
and
$$
\Epsig = \frac{(\epsilon^+ - \epsilon^-)^2}{12}
=
\frac{\left ( \Epbar \right )^2}{3}
$$
so that
$$
\frac{\Epbar}{\Epsig}
=
\frac{3}{\Epbar}
$$
Since only $\Epbar$ and a constant appear we can apply \equat{NSandEpbar} to obtain
\be
\label{eq:SUniform}
\boxed{
\Epbar = \left (\frac{\Ns}{\Ns(0)} \right )^{1/3}
}_{\,\,\, \mbox{Uniform}}
\ee

\vspace{0.125in}
\noindent{\bf \boldmath Gamma-Distributed ${\fee}(\epsilon)$:} 
Suppose 
$$
{\fee}(\epsilon)
=
\frac{\frac{k}{\Epbar_0}}{\Gamma(k)} (\frac{k}{\Epbar_0} \epsilon)^{k-1} e^{-{\frac{k}{\Epbar_0}\epsilon}}
$$
a Gamma distribution with parameter $k$ and mean $\Epbar_0$.  The variance is
$\frac{\Epbar_0^2}{k}$ so we have
$$
\frac{\Epbar}{\Epsig}
=
\frac{k}{\Epbar}
$$
which because $k$ is a fixed parameter allows us to use \equat{NSandEpbar} to obtain
\be
\label{eq:SGamma}
\boxed{
\Epbar = \left (\frac{\Ns}{\Ns(0)} \right )^{1/k}
}_{\,\,\,\mbox{Gamma} }
\ee
Note that increasing the order parameter $k$ decreases the dependence of $\Epbar$ on $\Ns$,
as we would expect since the distribution becomes more impulsive as $k$ grows.

\vspace{0.125in}
\noindent{\bf \boldmath Pareto ${\fee}(\epsilon)$:}
Suppose ${\fee}(\epsilon)$ is a Pareto distribution 
$$
{\fee}(\epsilon)
=
\alpha \epz^\alpha \epsilon^{-(1 + \alpha)}
$$
where $\epsilon \ge \epz > 0$ and $\alpha > 2$.  We have
$$
\Epbar = \frac{\alpha \epz}{\alpha -1}
$$
and
$$
\Epsig = \frac{\epz^2 \alpha}{(\alpha-1)^2(\alpha - 2)}
=
\frac{\Epbar^2 }{\alpha(\alpha - 2)}
$$
It is certainly tempting to follow the same route as the other examples --  divide $\Epbar$ by
$\Epsig$ and integrate using \equat{NSandEpbar}. However in this case, the parameter $\alpha$
depends on $\Epbar$ as in
$$
\alpha = \frac{\Epbar}{\Epbar -\epz}
$$

So, doing the requisite substitution we have
$$
\frac{\Epsig}{\Epbar}
=
\frac{(\Epbar - \epz)^2}{2 \epz - \Epbar}
$$

Integrating $\frac{\Epbar}{\Epsig}$ with respect to $\Epbar$ yields
$$
\int_1^\Epbar
\frac{2 \epz - \Epbar}{(\Epbar - \epz)^2}
d \Epbar
=
\frac{-\epz}{\Epbar - \epz} - \log ( \Epbar - \epz)
+
\frac{\epz}{1 - \epz} + \log ( 1 - \epz)
$$
which reduces to
\be
\label{eq:SPareto}
\boxed{
\log \left (\frac{\Ns}{\Ns(0)} \right )
=
\frac{\epz(\Epbar - 1)}{(\Epbar - \epz)(1 - \epz)} - \log \frac{\Epbar - \epz}{1 - \epz}
}_{\,\,\,\mbox{Pareto}}
\ee
Expressing $\Epbar$ compactly in terms of $\Ns$ is impossible so we are stymied in evaluating the
power relationship between $\Ns$ and $\Epbar(\Ns)$.  To establish that relationship requires new
machinery.

\subsection{\boldmath The Order Parameter $\order$}
\label{sect:order}
In the previous section we showed that $\Epbar$ can depend on $\Ns$, a key state variable in the
differential equations that govern contagion. Of particular note, we were able to show that if the
susceptibility is initially Gamma-distributed with shape parameter $k$, then
$\Epbar = (\Ns/\Ns(0))^{1/k}$ and that this relationship is maintained under the action of contagion
(see \equat{fgamma}). However, we know that the general action of contagion not only lowers $\Epbar$
over time but also changes the distribution shape as well. Thus, even if a closed form expression
for order can be obtained using \equat{NSandEpbar} with an initial susceptibility distribution
$g_0(\epsilon)$, then with the exception of Gamma distributions, the passage of time will change the
order. 

For instance, starting from an initially uniform distribution with order $1/3$ as determined in
\equat{SUniform}, \equat{general} will immediately produce a truncated exponential distribution
which over time will be substantively indistinguishable from a true exponential distribution.  Thus,
the initial order parameter would evolve from $1/3$ in \equat{SUniform} to $1$ (corresponding to a
Gamma distribution with shape parameter $k=1$).

We have already seen that $\Epbar$ may be a relatively complicated function of
$\Ns$ (\equat{SPareto}) as opposed to a simple power law (\equat{SGamma}).  Furthermore, in some cases it may even be impossible to compose the integrand of \equat{NSandEpbar} explicitly in terms of $\Epbar$.


We circumvent these difficulties by defining the instantaneous order, $\order$, as
\be
\label{eq:order}
\boxed{
{\order} \equiv \frac{d (\log \Epbar)}{d (\log \Ns)}
}
\ee
That is, the variation of the $(\log \Epbar)$ with $(\log \Ns)$ is explicitly a power law
relationship. And while certainly the slope defined by \equat{order} may change for different values
of $\Epbar$ and $\Ns$, it still defines a power law relationship between $\Epbar$ and $\Ns$ at a
given instant.

\Equat{order} provides a basis for investigating the range of power laws possible between $\Epbar$
and $\Ns$.  We can derive a lower bound on ${\order}$ by noting that both $\frac{d \Epbar}{dt}$ and
$\frac{d \Ns}{dt}$ are non-positive. Thus the ratio of their differentials in \equat{order} must be
greater than or equal to zero so that $\order \ge 0$.  We then note that via \equat{dEdt} we have
${\order = 0}$ in only two circumstances -- the contagion has run its course and $\dNIt = 0$, or the
variance of the susceptibility distribution is zero implying that all individuals have identical
susceptibilities.  For active contagion ($\dNIt >0$) this fact is worth memorializing:
\be
\label{eq:ordernonnegative}
\boxed{
\begin{array}{c}
\mbox{If $\dNIt > 0$, then $\order(t) \ge 0$}\\ \\
\mbox{with equality iff $\fe{t}(\epsilon)$ is singular.}
\end{array}
}
\ee
To summarize, $\Epbar(t)$ is {\em always} a function of the contagion state variable $\Ns$  unless all individuals have the same susceptibility or the contagion has run its course. Otherwise at any time $t$, $\Epbar(t) \propto \Nsorder$ where $\order(t) > 0$.

\subsection{\boldmath $\order$  and $\dorder$ in Terms of Moments}
Knowing $\order$ can only be zero or positive is useful. However, explicit evaluation of \equat{NSandEpbar}
to obtain $\order$ can be difficult -- witness the Pareto distribution considered previously.  Nonetheless, we can always calculate $\order$ (and even its time derivative $\dorder$)
in terms of the moments of $\fe{t}(\epsilon)$, either analytically or empirically.

To begin, we first write \equat{NSandEpbar} as
\be
\label{eq:W}
\log \left (\frac{\Ns}{\Ns(0)} \right )
=
\int_1^{\Epbar} \frac{\Epbar}{\Epsig} d \Epbar
=
{\Kw}(\log \Epbar)
\ee
We then have
$$
(d \log \Ns)
=
(d \log \Epbar)  {\Kw}^\prime (\log \Epbar)
$$
which via \equat{order} leads to
\be
\label{eq:inverseorder}
\frac{1}{\order}
=
{\Kw}^\prime (\log \Epbar)
\ee
where 
$$
{\Kw}^\prime (x) = \frac{d}{dx} {\Kw}(x)
$$
Then, we note that if 
$$
F(x) = \int f(x) dx
$$
we have
$$
\frac{d}{dt} F(x) = \dot{x} f(x)
$$
where "dot" implies differentiation with respect to time.
So, differentiating the two rightmost terms of \equat{W} yields
$$
\dEpbar \frac{\Epbar}{\Epsig}
=
\frac{\dEpbar}{\Epbar} {\Kw}^\prime (\log \Epbar)
$$
so that
$$
{\Kw}^\prime (\log \Epbar)
=
\frac{\Epbar^2}{\Epsig}
$$
and
\be
\label{eq:ordervstime}
\boxed{
\order
=
\frac{ \Epsig}{\Epbar^2}
=
\frac{\Eppbar}{\Epbar^2} - 1
}
\ee
where 
$$
\Eppbar = \int_0^\infty \epsilon^2 \fe{t}(\epsilon) d \epsilon
$$
We note that $\order$ as defined in \equat{ordervstime} is exactly the square of a quantity often
called the "coefficient-of-variation" \cite{Gomes2020}.

To determine how rapidly $\order$ changes it is most convenient to 
differentiate \equat{inverseorder} (as opposed to \equat{ordervstime}) to obtain
$$
\frac{d}{dt} \left ( \frac{1}{\order} \right )
=
\frac{\dEpbar}{\Epbar} {\Kw}^{\prime\prime} (\log \Epbar)
=
 \frac{2 \Epbar \dEpbar}{\Epsig}
-
\frac{\Epbar^2}{\Epsig^2} \dEpsig
$$

Remembering that $\Epsig= \Eppbar - \Epbar^2$
and ${\dEpbar} = -\beta{{\dNIt}}{{\Epsig}}$ we have
$$
\dEpsig
=
\frac{\dEpbar}{\Epsig}
(\Epppbar- \Eppbar \Epbar - 2 \Epbar \Epsig)
=
\frac{\dEpbar}{\Epsig}
E \left [ ({\cal E} - \Epbar)^3 \right ]
$$
If we then define the distribution "skew" as 
$$
\Epskew
\equiv
\frac{E \left [ ({\cal E} - \Epbar)^3 \right ]}{\sigma_{\cal E}^3}
$$
we have
\be
\label{eq:dorderinversedt}
\frac{d}{dt}
\left ( \frac{1}{\order} \right )
=
\frac{\dEpbar\Epbar}{\Epsig}
\left (
2 
-
\frac{\Epbar}{\sigma_{\cal E}} 
\Epskew
\right )
\ee
and since $\frac{d}{dt}
\left ( \frac{1}{\order} \right ) = - \frac{{\dorder}}{\order^2}$ we obtain
\be
\label{eq:dorderdt}
\boxed{
{\dorder}
=
\dEpbar \frac{\Epsig}{\Epbar^3}
\left (
\frac{\Epbar}{\sigma_{\cal E}} 
\Epskew
-2 \right )
}
\ee
Applying \equat{dEdt} yields
\be
\label{eq:dorderdt2}
\boxed{
{\dorder}
=
- \beta \Ni 
\frac{\sigma_{{\cal E}}^4}{\Epbar^3}
\left (
\frac{\Epbar}{\sigma_{\cal E}} 
\Epskew
-2 \right )
}
\ee
Therefore if the skew of a distribution is zero or negative, the order $\order$ under the action of
contagion would initially increase.  Alternatively, if there is strong positive skew so that
$\frac{\Epbar}{\sigma_{\cal E}} \Epskew -2 >0$ (as there would be for heavier-tailed distributions), then the order would initially decrease.

We can also define $\order$ (and $\dorder$, if desired) in terms of the Laplace transform of the initial distribution $g_0(\epsilon)$.  Defining the Laplace transform of $g_0(\epsilon)$ as
$$
G_0(s) = \int_0^\infty g(\epsilon)e^{-s\epsilon}d\epsilon
$$
we know that since $g_0(\epsilon)$ is a probability function we have
$$
\Epbar(0) = - G^\prime(0)
$$
and
$$
\Eppbar(0) = G_0^{\prime\prime}(0)
$$
So, via \equat{general} we have
$$
\Epbar(\beta \NIt)
=
- \frac{G_0^\prime (\beta \NIt)}{G_0(\beta \NIt)}
$$
and
$$
\Eppbar(\beta \NIt)
=
\frac{G_0^{\prime\prime}(\beta \NIt)}{G_0(\beta \NIt)}
$$
so that
\be
\label{eq:ordervialaplace}
\boxed{
\order(\beta\NIt)
=
\frac{G_0^{\prime\prime}(\beta \NIt)G_0(\beta \NIt)}{\left ( G_0^\prime (\beta \NIt)\right )^2} - 1
}
\ee
This approach is convenient because it requires one integration to find the Laplace transform of $g_0(\epsilon)$ and then only differentiations thereafter.

We must emphasize that while both \equat{ordervstime} and \equat{dorderdt} can be used to determine
snapshots of what the current order is and where it will go next, if the time courses of mean and
variance can be calculated for a given initial distribution $g_0(\epsilon)$, then the complete time
course of order $\order$ is known through \equat{ordervstime}.  Likewise, if the Laplace transform
of $g_0(\epsilon)$ is known, $\order$ can also be calculated through \equat{ordervialaplace}.  We
exercise these results in the next section.

\subsection{\boldmath Effective Order $\order$ Examples}
\vspace{0.125in}
\noindent{\bf \boldmath $2$-Point Distribution:}
\Equat{2point} via \equat{general} and \equat{ordervstime} yields
\be
\label{eq:K2point}
\boxed{
\order_{\mbox{\tiny $2$-Point}}
=
\left ( \frac{p}{1-p} \right ) e^{\frac{\beta \NIt \Epbar}{1-p}}
}
\ee
which starts at $\frac{p}{1-p}$ (agreeing with \equat{S2point}) and increases exponentially with $\beta \NIt$.

\vspace{0.125in}
\noindent{\bf Gamma Distribution:}
From \equat{ggamma}, the variance of a Gamma distribution is $\Epbar^2/k$ and the skew is
$2/\sqrt{k}$.  Evaluation of \equat{ordervstime} yields $\order = 1/k$ as expected from
\equat{SGamma}.  Likewise, evaluation of \equat{dorderdt} yields identically $0$, since the order is
always $1/k$ for a Gamma distribution with parameter $k$.  Thus
\be
\label{eq:KGamma}
\boxed{
\order_{\mbox{\tiny Gamma}}
=
\frac{1}{k}
}
\ee

\vspace{0.125in}
\noindent{\bf Uniform Distribution:}
We have via \equat{epbaruniform}
$$
\Epbar_{\mbox{\tiny Uniform}} (t)
=
\frac{1}{\beta \NIt} \cdot \frac{1 - \left( 1 + \beta \NIt \epmax\right) \cdot e^{-\beta \NIt \epmax}}{1 - e^{-\beta \NIt \epmax}}
$$
and using \equat{funiform} we calculate
$$
\Eppbar_{\mbox{\tiny Uniform}} (t)
=
\frac{2 - \frac{\beta\NIt \epmax e^{-\beta \NIt \epmax}(2+\beta \NIt \epmax)}{1 - e^{-\beta \NIt \epmax}}}{(\beta \NIt)^2}
$$
so that letting $\phi =\beta \NIt \epmax$ we have
\be
\label{eq:KUniform}
\boxed{
\order_{\mbox{\tiny Uniform}}
=
\frac{\left ( 1 -e^{-\phi} \right )\left ( 2 - (2 + \phi(2 + \phi))e^{-\phi} \right )}
{\left ( 1 - (1 + \phi) e^{-\phi} \right )^2}
-
1
}
\ee
We can can also calculate $\order$ for the uniform distribution using \equat{ordervialaplace}. 
The Laplace transform  of a uniform distribution on $[0, \epmax]$ is
$$
G_0(s) = \frac{1- e^{-s \epmax}}{s\epmax}
$$ 
so that
$$
- G_0^\prime (s) = \frac{1- e^{-s \epmax}}{s^2\epmax}
-\frac{e^{-s \epmax}}{s}
=
\frac{1 - (1 + s\epmax)e^{-s \epmax}}{s^2 \epmax} 
$$
and
$$
G_0^{\prime\prime}(s) = 
2\frac{1- e^{-s \epmax}}{s^3\epmax}
-\frac{e^{-s \epmax}}{s^2}
-\frac{e^{-s \epmax}}{s^2}
- \epmax \frac{e^{-s \epmax}}{s}
$$
which reduces to
$$
G_0^{\prime\prime}(s) = 
\frac{2 - (1 + (1 + s \epmax)^2)e^{-s \epmax}}{s^3\epmax}
$$
so that with $s \epmax = \beta \NIt \epmax = \phi$ becomes
via \equat{ordervialaplace}
\be
\label{eq:KUniformLaplace}
\boxed{
\order_{\mbox{\tiny Uniform (Laplace)}}
=
\frac{(1- e^{-\phi})(2 - (1 + (1 + \phi)^2)e^{-\phi})
}{\left ( 1 - (1 + \phi)e^{-\phi}  \right )^2}
}
\ee
\Equat{KUniformLaplace} is identical to \equat{KUniform}.

\vspace{0.125in}
\noindent{\bf Pareto Distribution:}
Now recall that we could not derive an explicit order for the Pareto distribution.  However, we know the time course of the distribution under contagion (\equat{fpareto}),
$$
\fe{t}(\epsilon)
=
\frac{\alpha \epz^\alpha  \epsilon^{-(1+\alpha)}}
{\alpha E_{(1+\alpha)}(\beta {\NIt} \epz)} e^{- \beta {\NIt} \epsilon}
$$
and we know the time course of the mean
$$
\Epbar_{\mbox{\tiny Pareto}} (t)
=
\epz \frac{E_{(\alpha)}(  \NIt \beta \epz)}{E_{(1 + \alpha)}( 
 \NIt \beta \epz)}
$$
and also $\Eppbar(t)$
$$
\Eppbar_{\mbox{\tiny Pareto}} (t)
=
\epz^2 \frac{E_{(\alpha-1)}(  \NIt \beta \epz)}
{E_{(1 + \alpha)}(\NIt \beta \epz)}
$$
so that
\be
\label{eq:KPareto}
\boxed{
\order_{\mbox{\tiny Pareto}} 
=
\frac{E_{(\alpha+1)}(  \NIt \beta \epz)E_{(\alpha-1)}(  \NIt \beta \epz)}
{E_{(\alpha)}^2(  \NIt \beta \epz)}
-
1
}
\ee
\begin{figure}
\noindent{\bf \boldmath \Large $\order$}
\begin{center}
\includegraphics[width=3.25in]{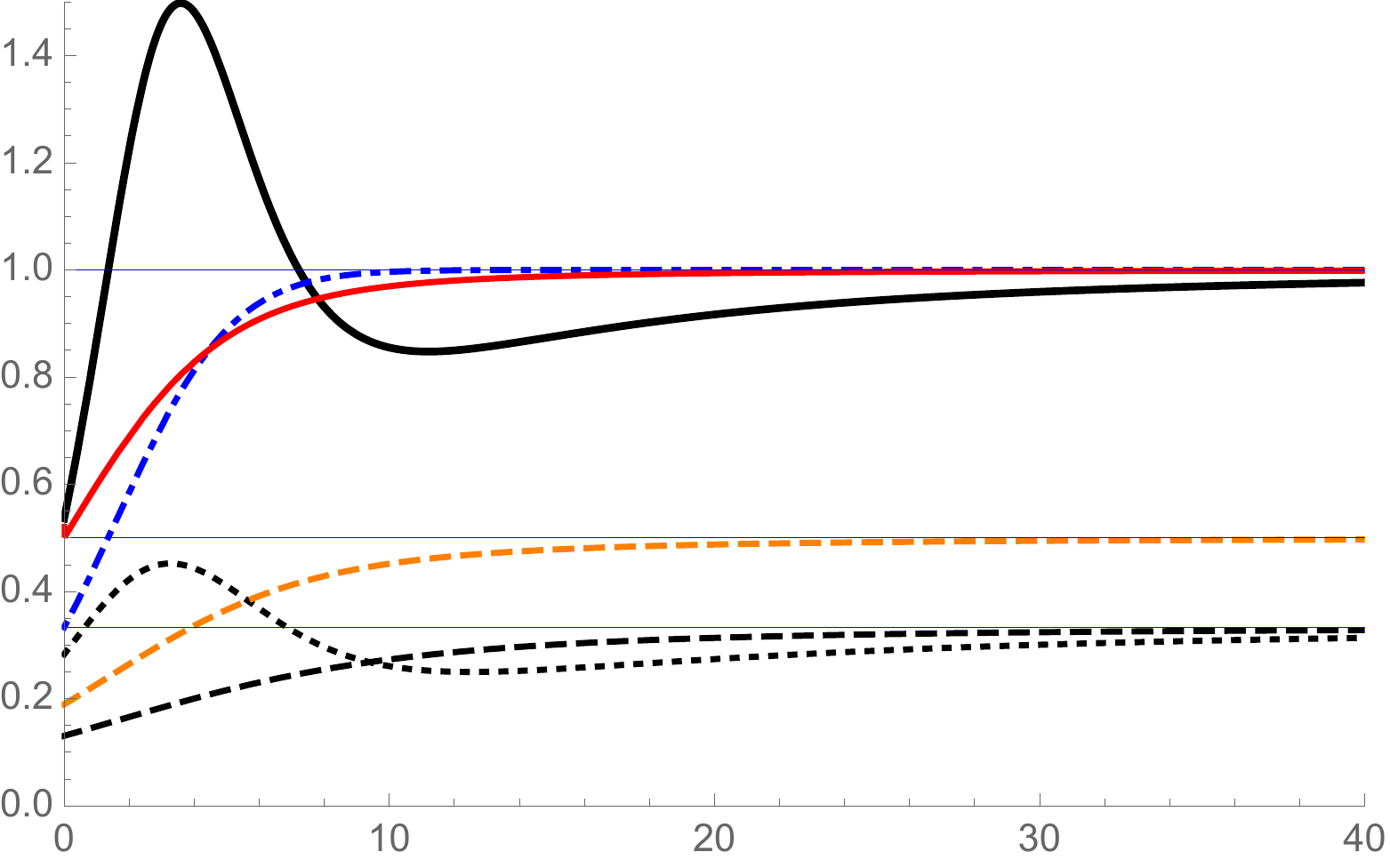}\\
{\bf \boldmath \Large $\phi$}\\
\end{center}
\vspace{-0.15in}
\caption{{\bf \boldmath $\order$ versus $\phi = \beta \NIt$:}
  Solid -- $g_0(\epsilon) = 1 + \cos 2 \pi \epsilon$;
  Dashed -- $g_0(\epsilon) = 1 - \cos 2 \pi \epsilon$;
  Dotted -- $g_0(\epsilon) = 1 - \cos 4 \pi \epsilon$;
  Dot-Dashed -- $g_0(\epsilon)$ uniform; 
  Red -- $g_0(\epsilon)$ downsloped linear;
  Orange -- $g_0(\epsilon) = \frac{\pi}{2} \sin \pi \epsilon$.  Note that the asymptotes comport with the orders of the limiting $k=1,2,3$ order Gamma distributions as seen in FIGURE~\ref{fig:FGang}{(d)}.}
\label{fig:Kgang}
\end{figure}
In FIGURE~\ref{fig:Kgang} we show the $\order$ evolution corresponding to the susceptibility
distribution evolution snapshots provided in FIGURES~\ref{fig:FGang}.
It is interesting to note
that while the order asymptotes comport with the convergence to Gamma distributions of orders
$k= 1,2,3$ seen in FIGURE~\ref{fig:Kgang}, the intermediate order (and thus the contagion dynamics)
may vary significantly from start to finish, depending upon the initial distribution, $g_0(\epsilon)$.

\section{The {\Boe}}
Revisiting \equat{dNsdt} and \equat{dNidt}, we see that the number of infections $\Ni$ starts to
wane when
$$
\beta \Epbar \Ns = \gamma
$$
That is, since both $\Epbar$ and $\Ns$ are strictly monotone decreasing functions of time, if
$\beta \Epbar(0) \Ns(0) > \gamma$, there is a single point $t^*>0$ at which $\dNit = 0$ and
\be
\label{eq:boe}
\Nsstar = \Ns(t^*) = \frac{\gamma}{\beta} \frac{1}{\Epbar(t^*)}
\ee
$1 - {\Nsstar}$ is defined as the {\em \boe}.  It should be noted that owing to the temporal
variation of $\Epbar(t)$, the usual final value results \cite{Ma2006} do not directly apply.  Thus,
$\Nsstar$ marks the beginning-of-the-end rather than {\em the} end of the contagion's course. 

Since we assume $\Epbar(0) =1$, we have $\Epbar(t) \le 1$ for $t > 0$, with equality iff the
susceptibility distribution is singular.  Thus, $\Nsstar$ is minimized iff the susceptibility
distribution is singular. We summarize this result as
\be
\label{eq:singularworst}
\boxed{
\Nsstar > \Nsstar_{\mbox{\tiny singular}} = \frac{\gamma}{\beta}
}
\ee
That is,  {\em a singular susceptibility distribution requires the largest proportion of individuals to be infected before contagion starts to wane.}

We could determine the {\boe} by brute force (numerical integration of \equat{dNsdt} and
\equat{dNidt}).  However, it is also possible to entirely avoid differential equation integration
and concomitant numerical errors. As previously, we define the Laplace transform of $g_0(\epsilon)$
as
$$
G_0(s) = \int_{0}^\infty  g_0(\epsilon)e^{-\epsilon s} d\epsilon
$$
By setting $\phi = \beta \NIt$, we can use \equat{general} to write
\be
\label{eq:epbarlaplace2}
\Epbar(\phi)
=
\frac{- G_0^\prime(\phi)}{G_0(\phi)}
\ee
and
$$
\Epsig(\phi)
=
\frac{G_0(\phi) G_0^{\prime\prime}(\phi) - (G_0^\prime(\phi))^2}{G_0^2(\phi)}
$$
so that
$$
\frac{\Epbar(\phi)}{\Epsig(\phi)}
=
\frac{-G_0^\prime(\phi) G_0(\phi)}{G_0(\phi) G_0^{\prime\prime}(\phi) - (G_0^\prime(\phi))^2}
$$

Then consider that by differentiating \equat{epbarlaplace2} we obtain
$$
\frac{d \Epbar(\phi)}{d\phi}
 =
- \frac{G_0(\phi) G_0^{\prime\prime}(\phi) - (G_0^\prime(\phi))^2}{G_0^2(\phi)}
$$
so that
$$
\label{eq:depbards2}
d \Epbar(\phi)
 =
- \frac{G_0(\phi) G_0^{\prime\prime}(\phi) - (G_0^\prime(\phi))^2}{G_0^2(\phi)} d\phi
$$
and the integrand $\frac{\Epbar(\phi)}{\Epsig(\phi)}  d \Epbar(\phi)$ in \equat{NSandEpbar} becomes
$$
\left ( \! \frac{-G_0^\prime(\phi) G_0(\phi)}{G_0(\phi) G_0^{\prime\prime}(\phi) \!- \!(G_0^\prime(\phi))^2} \! \right )
\!\!
\left (\! - \frac{G_0(\phi) G_0^{\prime\prime}(\phi) \!-\! (G_0^\prime(\phi))^2}{G_0^2(\phi)} \! \right ) d\phi
$$
which simplifies to
$$
\frac{G_0^\prime(\phi)}{G_0(\phi)} d\phi
$$
and allows us write
\be
\label{eq:intermediateNS2}
\log \left ( \frac{\Ns}{\Ns(0)} \right )
=
\int_0^{\phi_{\Epbar}}  \frac{G_0^\prime(\phi)}{G_0(\phi)} d\phi
=
\log G_0(\phi) \bigg\rvert_0^{\phi_{\Epbar}}
\ee
where $\phi_{\Epbar}$ is the value of $\phi$ for which \equat{epbarlaplace2} evaluates to $\Epbar$.  Since $G_0(0)= 1$, \equat{intermediateNS2} reduces to
\be
\label{eq:NSandEpbarLaplace2}
\Ns (\Epbar(\phi_{\Epbar}))
=
G_0(\phi_{\Epbar}) \Ns(0)
\ee
Then, since $\Ns(\cdot)$ is effectively parametrized in $\phi$, we can 
rewrite \equat{NSandEpbarLaplace2} as
$$
\Ns(\phi)
=
G_0(\phi)  \Ns(0)
$$
and use \equat{epbarlaplace2} to obtain $\Epbar(\phi)$ so that we have
$$
\Epbar(\phi) \Ns(\phi) = \frac{- G_0^\prime (\phi)}{G_0(\phi)} G_0(\phi) \Ns(0) = - G_0^\prime(\phi)
\Ns(0)
$$
We can then identify the value $\phi^*$ for which \equat{boe} is satisfied via
\be
\label{eq:Gboe}
\boxed{
-G_0^\prime(\phi^*) \Ns(0) = \frac{\gamma}{\beta}
}
\ee
to obtain the {\boe} as
\be
\label{eq:xsboe}
\boxed{
1 - \Nsstar = 1- G_0(\phi^*)\Ns(0)
}
\ee

We can now exercise \equat{Gboe} and \equat{xsboe} for different $g_0(\epsilon)$.  However, since
$-G_0^\prime(\phi)$ is strictly monotone decreasing we must always assume
\be
\label{eq:G0LB}
-G_0^\prime(0) \Ns(0)
=
\Epbar(0) \Ns(0)
\ge
\frac{\gamma}{\beta}
\ee
Otherwise the solution $\phi^*$ to \equat{Gboe} does not exist.  Put another way, if \equat{G0LB} is
violated, contagion fizzles out.

\vspace{0.125in}
\noindent {\bf Singular Distribution:}
We have
$$
G_0(\phi)
=
e^{-\phi}
$$
and
$$
-G_0^\prime(\phi)
=
e^{-\phi}
$$
so that 
$$
\phi^* = -\log \frac{\gamma}{\beta \Ns(0)}
$$
and 
\be
\label{eq:herdsingular}
\boxed{
\Nsstar
=
\frac{\gamma}{\beta}}_{\,\,\, \mbox{\footnotesize herd singular}}
\ee
which yields $\Nsstar = 0.5$ if $\frac{\gamma}{\beta} = \frac{1}{2}$.

\vspace{0.125in}
\noindent {\bf \boldmath $2$-Point Distribution:}
We have
$$
G_0(\phi) = p + (1-p) e^{-\frac{1}{1-p} \phi}
$$
so that
$$
-G_0^\prime (\phi) = e^{-\frac{1}{1-p} \phi}
$$
We then have
$$
\phi^* = -(1-p) \log \frac{\gamma}{\beta\Ns(0)}
$$
so that
\be
\label{eq:herd2point}
\boxed{
\Nsstar=
p\Ns(0) + (1-p) \frac{\gamma}{\beta}}_{\,\,\, \mbox{\footnotesize herd 2-Point}}
\ee
with the proviso that $\Ns(0) \ge \frac{\gamma}{\beta}$ so that $\Ns(0) - \Nsstar \ge 0$.  This
restriction comports with the fact that the worst herd immunity threshold is $\frac{\gamma}{\beta}$
as given in \equat{herdsingular}.  If $\frac{\gamma}{\beta} = \frac{1}{2}$ and $\Ns(0) = 1$ we then
have
$$
\Nsstar
=
\frac{1}{2}(1+p)
$$

\vspace{0.125in}
\noindent {\bf Uniform Distribution:}
We have
$$
G_0(\phi) = \frac{1 - e^{-2 \phi}}{2\phi}
$$
and
$$
-G_0^\prime(\phi)
=
\frac{1 - e^{-2\phi}}{2\phi^2}
-\frac{2e^{-2\phi}}{2\phi}
$$
If $\frac{\gamma}{\beta} =\frac{1}{2}$ and $\Ns(0) = 1$, we numerically find $$
\phi^* =0.546
$$
and thence
$$
\Nsstar
=
0.609
$$

\vspace{0.125in}
\noindent {\bf Gamma Distribution:}
We have
$$
g_0(\epsilon)
=
\frac{k}{\Gamma(k)} (k \epsilon )^{k-1} e^{-k\epsilon}
$$
and
$$
G_0(\phi)
=
\left (
\frac{1}{\frac{\phi}{k} + 1}
\right )^k
$$
so that 
$$
G_0^\prime (\phi)
=
- \left (
\frac{1}{\frac{\phi}{k} + 1}
\right )^{k+1}
$$
so that
$$
\phi^*
=
k\left ( 
\left ( 
\frac{\gamma}{\beta}
\right )^{-\frac{1}{k+1}}
-1 \right )
$$
and
\be
\label{eq:herdgamma}
\boxed{
\Nsstar
=
\left ( \Ns(0) \right )^{\frac{1}{k+1}}
\left ( 
\frac{\gamma}{\beta}
\right )^{\frac{k}{k+1}}}_{\,\,\, \mbox{\footnotesize herd Gamma}}
\ee
Then, for $\frac{\gamma}{\beta} = \frac{1}{2}$, $\Ns(0) = 1$ and 
$$
k = \{0.5, 1.0, 2.0\}
$$ 
we have
$$
\Nsstar = \{0.794, 0.707, 0.630\}
$$

\vspace{0.125in}
\noindent {\bf Pareto Distribution:}
We have
$$
\epsilon_0 \frac{\alpha}{\alpha - 1}
=
1
$$
so that
$$
\epsilon_0
=
\frac{\alpha-1}{\alpha}
$$
and thence
$$
g_0(\epsilon)
=
\alpha 
\left ( \frac{\alpha-1}{\alpha} \right )^\alpha
\epsilon^{-(\alpha + 1)}
$$
so that
$$
G_0(\phi)
=
\alpha
E_{1+\alpha} \left (\frac{\alpha-1}{\alpha} \phi \right )
$$
and
$$
-G_0^\prime(\phi)
=
(\alpha - 1)
E_{\alpha}\left (\frac{\alpha-1}{\alpha} \phi \right )
$$
If $\frac{\gamma}{\beta} = \frac{1}{2}$, $\Ns(0) = 1$ and 
$$
\alpha = \{1.1, 1.5, 2.0, 3.0\}
$$
we have
$$
\phi^* = \{0.00554,0.367,0.535, 0.629 \}
$$
and
$$
\Nsstar = \{0.997,0.886 ,0.887, 0.828 \}
$$

\section{Discussion}
We have shown that the shape of the population susceptibility distribution can significantly affect
the time course of contagion and its ultimate severity. Since contagion modeling must ultimately
be in the service of contagion understanding and control, two issues immediately come to mind:
\begin{itemize}
\item Given an initial population susceptibility density $g_0(\epsilon)$, might there be good
  targeted intervention strategies for contagion control?
\item If population susceptibility is indeed variable, how might we efficiently and rapidly measure
  $\fe{t}(\epsilon)$?
\end{itemize}
We discuss these issues in the next two subsections.

\subsection{Intervention}
Suppose we are allowed to intervene and change some fraction of population susceptibilities.  What
reassignment maximizes the resulting {\boe}?  The intuitively obvious answer is to inoculate that
fraction of individuals, effectively setting their susceptibilities to zero.  Likewise, if the
particular fraction of the population can be chosen it seems equally obvious that we should choose
those individuals with greatest susceptibility.  

It should be noted, however, that implementing the susceptibility zeroing abstraction faithfully may
be difficult depending upon the practical methods available to mute susceptibility.  For instance,
perfect protection (through inoculation, isolation, and/or behavior modification) of individuals
serving critical high-exposure societal functions may be impossible. Furthermore, even if those
individuals who take ill are effectively removed from the equation, others must take their place,
which may result in no change to the population susceptibility distribution.  Nonetheless, assuming
we {\em could} sculpt the population susceptibility distribution through intervention, it is still
useful to show analytically that zeroing individual susceptibilities produces the best {\boe}, and
that the absolute best {\boe} is achieved by inoculating that fraction of individuals with the
highest susceptibilities.

To begin, let $g_0(\epsilon)$, the initial susceptibility distribution, be the weighted sum
of two arbitrary  singularity-free distributions $g_1(\epsilon)$ and $g_2(\epsilon)$:
\be
\label{eq:g0def}
g_0(\epsilon)
=
(1-p) g_1(\epsilon) + p g_2(\epsilon)
\ee
where $0< p < 1$.

$\Nsstar$ is obtained through \equat{xsboe} as
\be
\label{eq:herd}
\frac{\Nsstar}{\Ns(0)} = G_0(\phi^*) = (1-p) G_1(\phi^*) + p G_2(\phi^*)
\ee
where $\phi^*$ satisfies \equat{Gboe}:
\be
\label{eq:GPrime}
G_0^\prime (\phi^*) =  \left ( (1-p) G_1^\prime(\phi^*) + p G_2^\prime(\phi^*) \right )
=
-\frac{\gamma}{\Ns(0)\beta}
\ee
We then seek a replacement for $g_2(\epsilon)$ that maximizes $\Nsstar$.

Since $-G_1^\prime(\phi)$ is monotonically decreasing in $\phi$ we can minimize $\phi^*$ in
\equat{GPrime} by setting $G_2^{\prime}(\phi) = 0$ which implies
$g_2(\epsilon) = \delta (\epsilon)$.  Since $G_1(\phi)$ is also monotonically decreasing in $\phi$,
minimizing $\phi^*$ maximizes $G_1(\phi^*)$.  Then we note that setting
$g_2(\epsilon) = \delta(\epsilon)$ also produces maximum $G_2(\phi) = 1$ $\forall \phi$.  Therefore,
taking the probability mass $p$ associated with $p g_2(\epsilon)$ and relocating it to
$\epsilon = 0$ maximizes \equat{herd}.

Having established that we should inoculate the population fraction represented by
$p g_2(\epsilon)$, we can now consider how $g_1(\epsilon)$ should be chosen to absolutely maximize
$\Nsstar$ under the constraint of \equat{g0def}. Since it is always best to set
$g_2(\epsilon) = \delta(\epsilon)$, we rewrite \equat{herd} with $G_2(\phi) = 1$ as
\be
\label{eq:herd0}
\frac{\Nsstar}{\Ns(0)} = (1-p) G_1(\phi^*) + p
\ee
Then we consider that since
$$
G_1(\phi) = \int g_1(\epsilon) e^{-\epsilon \phi} d \epsilon
$$
and $e^{-\epsilon \phi}$ is strictly monotone decreasing in $\epsilon$, we can maximize $G_1(\phi)$
$\forall \phi > 0$ (and thereby \equat{herd0}) by placing as much probability mass as possible `` on
the left'' in $ \epsilon \in [0, \epsilon^*]$ with $\epsilon^*$ chosen to satisfy
\be
\label{eq:epsilonstar}
\int_0^{\epsilon^*} g_0(\epsilon) d \epsilon = 1-p
\ee

\Equat{g0def} requires $(1-p)g_1(\epsilon) \le g_0(\epsilon)$ since probability densities cannot be
negative.  Thus, setting $g_1(\epsilon) = g_0(\epsilon)/(1-p)$ on $[0 < \epsilon \le \epsilon^*]$
and zero elsewhere moves the maximum allowable amount of probability mass to the left and thereby uniquely
maximizes $G_1(\phi)$ $\forall \phi$.  Applying this result to the definition of $g_0(\epsilon)$ in
\equat{g0def} leads to,
\be
\label{eq:intervention}
\begin{array}{c}
  g_1(\epsilon) =
  \twodef{g_0(\epsilon)}{\epsilon \le \epsilon^*}{0}{\mbox{o.w.}}\\
  g_2(\epsilon) =
  \twodef{g_0(\epsilon)}{\epsilon > \epsilon^*}{0}{\mbox{o.w.}}
\end{array}
\ee
That is, $g_1(\epsilon) = g_0(\epsilon)$ for $\epsilon \in [0,\epsilon^*]$ and zero elsewhere is the
``head'' of $g_0(\epsilon)$ and $g_2(\epsilon) = g_0(\epsilon)$ for
$\epsilon \in (\epsilon^*, \infty)$ and zero elsewhere is the ``tail'' of $g_0(\epsilon)$.

We note that if $g_0(\epsilon)$ contain singularities, then it may be impossible to satisfy
\equat{epsilonstar} as written. However, the same driving principle holds -- placing as much
probability mass as possible to the left in $g_1(\epsilon)$.  We would thus relax the strict
inequality in \equat{intervention} to allow some fraction of the singular mass at $\epsilon^*$ to
remain in the tail $g_2(\epsilon)$ such that
$(1-p)g_1(\epsilon^*) + p g_2(\epsilon^*) = g_0(\epsilon^*)$ while still satisfying
\equat{epsilonstar}.

So, as expected, if we can intervene during the progression of contagion and reassign some fraction
$p$ of susceptibilities, we should choose those individuals with greatest susceptibility and
inoculate them.  The result also suggests a simple inoculation strategy if we wish to immediately
quell contagion: inoculate a fraction $p$ sufficient to drive $\Epbar =
\frac{\gamma}{\beta}$. Assuming $\Epbar(0) = 1$ and $\frac{\gamma}{\beta} = \frac{1}{2}$ this means
we must inoculate $50\%$ of the population if everyone has the same susceptibility,
$\approx\!\!\! 30\%$ if the initial susceptibility distribution is uniform and $\approx\!\! 19\%$ if
the initial susceptibility distribution is exponential.

\subsection{Susceptibility Variation Measurement}
The notion of contagion intervention and control based on population susceptibility distribution begs the
question of how susceptibility \cite{Kwiatkowski2000} can be measured.  There are perhaps
immunological assays that could be applied to a population which could determine the likelihood that
a given individual would succumb to the illness after exposure to some unit dose.  Given
the difficulty and expense associated with timely testing for infection, such an approach may be
unwieldy.  Furthermore, if it is likely that individuals drawn from an immunologically naive
population have near identical innate dose/response reactions to a particular contagion, then not
only would such pre-infectious medical monitoring be costly, it would also be useless since
differentially applied interventions based on susceptibility would have no effect on contagion
progression.

However, if the specific contagion can only be transmitted through proximate contact (as opposed to
truly airborne over large distances), then two obvious measures of susceptibility come to mind:
\begin{itemize}
\item
  protective behaviors (e.g., mask-wearing and hygiene)
\item
  number of contacts \cite{Balcan2009,Lloyd-Smith2005}
\end{itemize}
Poor hygiene, lack of protection and high numbers of contacts all potentially result in higher
cumulative contagion dose and thus a higher probability of becoming infected. While hygiene
monitoring seems difficult (if not invasive), surveillance and telecommunications infrastructure,
suitably anonymized, might allow some measure of susceptibility variation to be obtained.
Mask-wearing volume could be measured and close contact recorded through cell phone records.  Of
particular interest, neither of these methods would rely on medical testing {\em a priori} so that
these proxies for susceptibility would {\em lead} as opposed to {\em lag} contagion and permit more
effective targeted contagion control.

Of course, whether contact intensity and observable behavior {\em are} reasonable proxies for
susceptibility is debatable.  Nonetheless, it seems worthwhile to examine whether it is possible to
cobble together at least a rough susceptibility profile estimate for a population that could inform
public health interventions -- again, ahead of as opposed to lagging contagion as all
medically-based detection necessarily does.

\section{Conclusion}
We have mathematically refined the insights first introduced in \cite{Peterson2020} to show how
population susceptibility variation under an assumption of static individual susceptibility affects
the dynamics of contagion progression.  Specifically, by positing that susceptibility might vary
over a population, we defined the population susceptibility probability density $\fe{t}(t)$, the
time-varying average susceptibility $\Epbar(t)$ and developed closed-form expressions to show how
these modifications to the usual SIR differential equations affect the dynamics of contagion and at
what population fraction we can expect herd immunity to begin muting it.  We showed that a
population with singular susceptibility (everyone has the same static susceptibility) has the worst
herd immunity threshold and the worst response to intervention in terms of what fraction of
individuals must be inoculated to initiate herd immunity.  We also showed that for a variety of
possible population susceptibility distribution assumptions that the herd immunity threshold could
be much lower and concomitantly, the effects of intervention more potent.

We then discussed population susceptibility measurement through the proxies of individual mobility
and contact intensity as well as individual protective behaviors (such as mask-wearing).  If these
are indeed reasonable and lag-less proxies for susceptibility, the use of non-medical electronic
susceptibility monitoring and the closed-form contagion state expressions derived here seems an
interesting line of research in the prediction and control of contagion.  Combined with recent
hypotheses suggesting population susceptibility variation changes the progression and ultimate
severity of SARS-CoV-2 \cite{Gomes2020,Britton2020,Dufresne2020}, we feel that real-time measurement
of susceptibility could be a critically important determinant of policy to control future pandemics.

\bibliographystyle{unsrt}
\bibliography{second-order,MERGE11}

\end{document}